# Characterization of specific nuclear reaction channels by deconvolution in the energy space of the total nuclear cross-section of protons - applications to proton therapy and technical problems (transmutations)

W. Ulmer

Medical Physics of the Radio-Oncology North-Wuerttemberg/Baden-Baden and MPI of Physics, Goettingen, Germany

**Abstract** - The total nuclear cross-section $Q^{tot}(E)$ resulting from the interaction of protons with nuclei is decomposed in 3 different contributions: 1. elastic scatter at the complete nucleus, which adopts a part of the proton kinetic energy; 2. inelastic scatter at a nucleus, which changes its quantum numbers by vibrations, rotations, transition to highly excited states; 3. proper nuclear reactions with change of the mass and/or charge number. Then different particles leave the hit nucleus (neutrons, protons, etc.), which is now referred to as 'heavy recoil' nucleus. The scatter parts of $Q^{tot}(E)$ according to points 1 and 2 can be removed by a deconvolution acting at $Q^{tot}(E)$ in the energy space. The typical nuclear reaction channels are mainly characterized by resonances of a reduced cross-section function $Q^{red}(E)$. The procedure is applied to cross-sections of therapeutic protons and also to $Cs_{55}^{137}$ as an example with technical relevance (transmutations with the goal to drastically reduce its half-time).

## Introduction

Owing to the increasing importance of the radiotherapy of protons sophisticated elaboration of nuclear interactions play a significant role in therapy planning. In particular, we think of the release of secondary protons, neutrons and fission products of nuclei leading to heavy recoils. In previous papers we have developed calculation methods for the determination of nuclear cross-sections by proton hits up to 270 MeV [1 - 4]. These methods are based on generalized nuclear shell theory in order to take account for interactions of protons with nuclei by various highly excited and by inclusion of virtual states. The task of a second step has been the transcription to collective models. Thus we have shown that this description can be carried out by three Gaussian distributions with shifts of the maximal values for the total cross-section $Q^{tot}(E)$ and an error function erf(E) for the asymptotic behavior [1]. A comparison study with the generalized Bethe-Wigner-Flügge formula for resonances and further data [5 - 8] has revealed that most processes leading to $Q^{tot}(E)$ are indeed inelastic scatter processes, since the total quantum numbers of the system are changed. However, there are also elastic contributions without changing of any quantum number resulting from the energy transfer of the proton to the whole nucleus yielding scatter deflections of the incident proton. According to the cross-section formula given in section 2, it is possible to separate the elastic contribution and that inelastic part of the interaction process leading only to a change of quantum numbers of a nucleus by a deconvolution in the energy space (the nuclear mass number $A_N$ and charge number Z remain unchanged). Elastic/inelastic scatter without changing $A_N$ and Z lead to the property of $Q^{tot}(E)$ that we can only verify one discrete maximum. The goal of this study is to characterize the proper nuclear interactions of protons due to changes of $A_N$ and Z of a nucleus by the deconvolution of $Q^{tot}(E)$. In therapy planning, the interaction of protons with carbon, oxygen and calcium are of importance, whereas the interaction with copper has only relevance for the behavior of the beam-line. Since all parameters of the investigations presented in this communication only depend on $A_N$ and Z, it appears to be straightforward to replace them in voxel-based CT information by the substitutions:

$A_N \rightarrow A_{N,effective}$ and $Z \rightarrow Z_{effective}$, which incorporate the essential starting-point for the inclusion of nuclear reactions in modern proton therapy planning algorithms. However, the presented method may not be restricted to proton therapy planning problems, and we shall also consider an application to the interaction of protons with $Cs_{55}^{137}$ implying a change of the long half-time of about 30 years of $\mathbf{Cs_{55}^{137}}$ to reaction products with a comparably rather short life-time. The procedure is referred to as *'transmutation of nuclei'* with the goal to circumvent the extremely long duration of the storage of nuclear decay products emerging from nuclear reactors.

A further purpose of the study is a calculation of the decrease of the fluence of primary protons and their increase of secondary particles in media like water-equivalent materials, bone (calcium) and copper (beam-line), because the transport of secondary particles is not sufficiently accounted for in therapy planning systems. In particular, the contributions of neutrons, deuterium, tritium and α-particles as secondary products need a sophisticated analysis. One should take into account that Figures 1 and 2 result from the integration of the total nuclear cross-section $Q^{tot}(E)$. This procedure has been worked out by Segrè [7] and applied in all calculations of the previous publications [2 - 4, 9]. Thus the difference between primary protons and secondary particles (energy spectrum, scatter angles) can be accounted for.

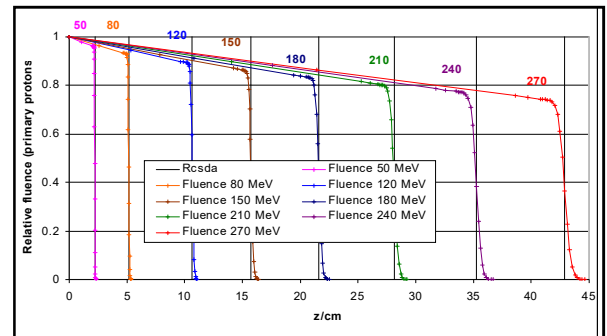

**Figure 1:** Decrease of proton fluence in water (see paper [1]).

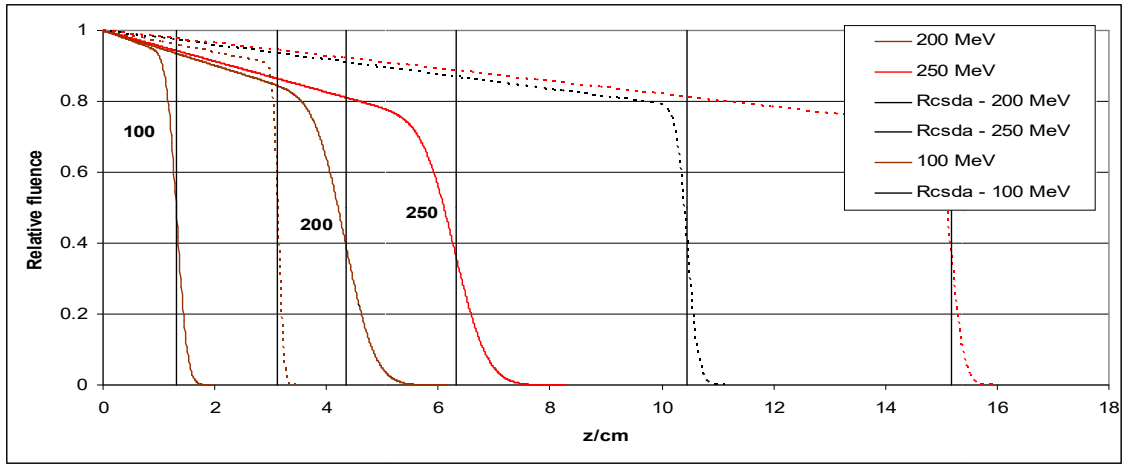

**Figure 2:** Decrease of the fluence of 100, 200, and 250 MeV primary protons in copper (solid lines) and in calcium (dashes). The $R_{CSDA}$ ranges are stated by perpendicular lines (see [1]).

## 1. Methods

With regard to all succeeding aspects it is important to look at Figure 3, which shows the nuclear potential with the range $R_{strong}$ of strong interactions. One should be aware of the property that with regard to the cross-section the concentration of nucleons within potential only amounts to ca. 4/9 of the square of the potential cross-section. This fact enables us to consider proper nuclear reactions denoted by point 3 (abstract) and expressed by equations (1, 1a) and, in addition, elastic/inelastic scatter at a nucleus (points 1 and 2 in the abstract). Therefore the following properties hold:

$$R_{strong} = 1.2 \cdot 10^{-13} \cdot \sqrt[3]{A_N} \; cm. \quad (1)$$
$$R_{Nucleus} \cong \tfrac{2}{3} \cdot R_{strong}. \quad (1a)$$

Thus formulae (1) and (1a) represent good approximations of spherical nuclei, i.e. $Z \approx A_N/2$ holds. It should be pointed out that the properties given by Figure 3 qualitatively hold for other nuclei.

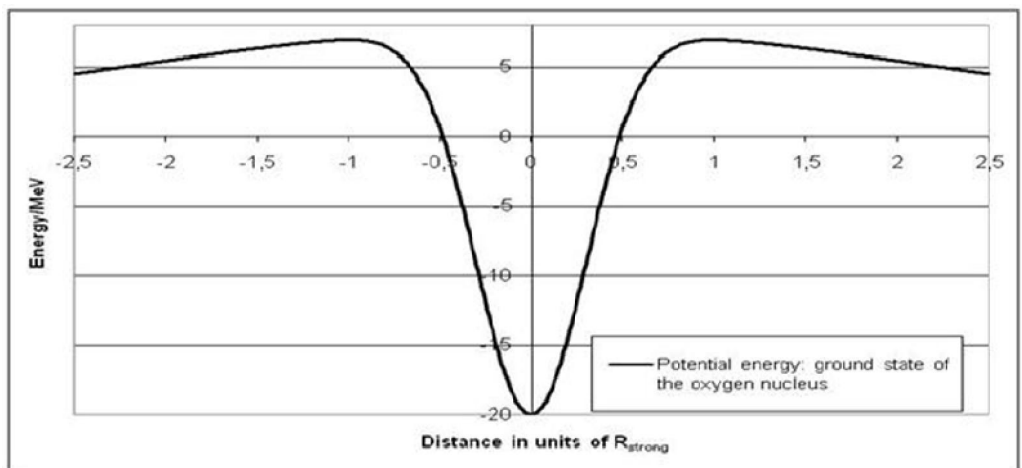

**Figure 3:** Nuclear potential of the oxygen nucleus in terms of units of $R_{strong}$.

Thus mainly the threshold energy $E_{Th}$ at $R = R_{strong}$ and the depth of the nuclear potential change by depending on Z and $A_N$. In order to perform deconvolutions of $Q^{tot}(E)$ with regard to a decomposition into $Q^{red}(E)$ and an *elastic/inelastic* scatter part, it is convenient to make use of the formulae described in a previous study [1], since all necessary calculations procedures can be handled analytically.

## 1.1. Previously derived calculation formulae of nuclear cross-sections using Gaussian distributions and the error function erf(E)

The total nuclear cross-section $Q^{tot}(E)$ of the proton - nucleus interaction is described by formula (2):

$$Q^{tot} = w_0 \cdot Q^{tot}_{as} \cdot erf(\frac{E-E_{Th}}{\sigma_{as}}) + w_{g0} \cdot \exp(-(\frac{E+\delta-E_{res}}{\sigma_{res}})^2) +$$
$$w_{g1} \cdot A_m \cdot \exp(-(\frac{E-E_{res}}{\sigma_1})^2) \; w_{g1} \cdot (1-A_m) \cdot \exp(-(\frac{E-\gamma-E_{res}}{\sigma_2})^2) + A_{boundary} \quad (2)$$

The parameters of formula (2) are given by formulas (2a, 3) and Tables 1 and 2.

$$\begin{aligned}
A_{boundary} &= (1-A_m) \cdot wg_0 - A_m \cdot wg_{00} + (1-A_m) \cdot (wg_0 - wg_{f_0}) \cdot erf(\frac{E-E_{Th}}{\sigma_{as}}) + \\
&\quad A_m \cdot erf(\frac{E-E_{Th}}{\sigma_{as}}) \cdot (wg_{00} - wg_{f_1}) + wg_{f_3} \cdot w_{Gauss} + \\
&\quad + w_{Gauss} \cdot wg_0 - A_m \cdot wg_{00} + (1-A_m) \cdot (wg_0 - wg_{f_0}) \cdot erf(\frac{E-E_{Th}}{\sigma_{as}}) + \\
&\quad A_m \cdot erf(\frac{E-E_{Th}}{\sigma_{as}}) \cdot (wg_{00} - wg_{f_1}) + wg_{f_3} \cdot w_{Gauss} + w_{Gauss} \cdot (wg_{f_3} - wg_{f_2}) \\
wg_0 &= \exp(-(\frac{E_{Th}-\gamma-E_{res}}{\sigma_2})^2); \; wg_{00} = \exp(-(\frac{E_{Th}-E_{res}}{\sigma_1})^2); \; wg_{f_0} = \exp(-(\frac{E_{res}-E_f-\gamma}{\sigma_2})^2; \; E_f = 270 \, MeV \\
wg_{f_1} &= \exp(-(\frac{E_{res}-E_f}{\sigma_1})^2); \; wg_{f_2} = \exp(-(\frac{E_{res}+\delta-E_f}{\sigma_{res}})^2); \; wg_{f_3} = \exp(-(\frac{E_{Th}-E_{res}+\delta}{\sigma_{res}})^2
\end{aligned} \quad (2a)$$

**Table 1:** Some essential data of the energy E and $Q^{tot}(E)$ necessary in formulas (2) and (2a)

| Nucleus | $E_{Th}$/MeV | $E_{res}$/MeV | $Q^{tot}_{max}$/mb | $Q^{tot}_c$/mb | $Q^{tot}_{as}$/mb |
|---|---|---|---|---|---|
| C  | 5.7433 | 17.5033 | 447.86  | 426.91  | 247.64  |
| O  | 6.9999 | 20.1202 | 541.06  | 517.31  | 299.79  |
| Ca | 7.7096 | 25.2128 | 984.86  | 954.82  | 552.56  |
| Cu | 8.2911 | 33.4733 | 1341.94 | 1308.07 | 752.03  |
| Cs | 8.5620 | 51.2022 | 2009.71 | 1726.48 | 1254.93 |

The parameters of Table 1 can be calculated by the parameters of Table 2 used in a function of Z and $A_N$ according to formula (3):

$$P_p = C_p \cdot Z^p / A_N^q . \quad (3)$$

**Table 2:** Parameters $P_p$ of the cross-section formula (3)

| Parameters $P_p$ in formula (3) | $C_p$ | p | q |
|---|---|---|---|
| $w_{Gauss}$ | 36.05 | 1.421 | 1.811 |
| $\delta$ | 0.09335 | -1.621 | -0.405 |
| $\gamma$ | -9.155 | 2.396 | 1.763 |
| $\sigma_{res}$ | 0.925 | -1.232 | -1.595 |
| $\sigma_1$ | 17.215 | 0.6375 | 0.31 |
| $\sigma_2$ | 11.575 | 1.13 | 0.38 |
| $\sigma_{as}$ | 1.074 | 1.745 | 2.102 |
| $A_m$ | 0.06257 | -1.102 | -1.335 |
| $E_{res}$ | 4.2064 | -0.7932 | 1.1561 |
| $w_{as}$ | 47.354 | 0.0055 | -0.6616 |

## 1.2. Deconvolution methods

Deconvolutions of cross-section formula (2) to determine $Q^{tot}(E)$ can be performed with regard to the elastic and to an inelastic part (change of the nuclear quantum numbers) of nuclear interactions (proton scatter), i.e. if $A_N$ and Z remain constant! This means that if a proton hits the domain of strong interaction a part of the energy is transferred to the whole nucleus. Thus it is possible that the elastic scatter is the only interaction, but usually beside this part other excitations can occur: vibrations by deformations of the nucleus, rotations, and nuclear excited states. The proper nuclear reactions result from extremely high excited and/or virtual states. We consider the general Gaussian convolution kernel:

$$K(\varepsilon, \xi - \xi') = \varepsilon^{-1} \pi^{-1/2} \cdot \exp(-(\xi-\xi')^2 / \varepsilon^2). \quad (4)$$

Usually ξ and ξ' are identified as co-ordinates in the position space, ρ(x) is the source function and φ(x) an image function, respectively, then the following integral transform is applied in many problems:

$$\varphi(x) = \int K(\varepsilon, u - x)\rho(u)du. \quad (4a)$$

Examples of equation (4a) are blurring of images or the influence of the finite detector size in proton/photon dose calculations [10], in this case the dimension of ε is a length. Formula (4a) exists, if ρ belongs to the function space $L_1$. The kernel (4) is the result of a Green's function of the operator function [10]:

$$O^{-1} = \exp(0.25 \cdot \varepsilon^2 \cdot d^2/d\xi^2). \quad (5)$$

This operator function is defined as a Lie expansion (for details, see appendix A). However, the application of $O^{-1}$ according to equation (4a) might be different from the integral operator notation, since ρ(x) now has to belong to $C^\infty$:

$$\varphi(x) = O^{-1} \cdot \rho(x). \quad (5a)$$

The inverse problem of relation (5a) results from multiplication with the operator $O^1$ on both sites to obtain:

$$\rho(x) = O^1 \cdot \varphi(x), \quad (6)$$

where $O^1$ is determined by:

$$O^1 = \exp(-0.25 \cdot \varepsilon^2 \cdot d^2/d\xi^2). \quad (6a)$$

The operators $O^1$ and $O^{-1}$ satisfy the relation

$$O^1 \cdot O^{-1} = O^{-1} \cdot O^1 = 1 \text{ (unit operator)}. \quad (6b)$$

The function class $C^\infty$ related to $O^1$ and $O^{-1}$ implies that for each function of this function space continuous derivations of arbitrary order have to exist; this fact is a consequence of corresponding Lie series expansions of $O^1$ and $O^{-1}$ (see appendix A). The integral operator notation of relation (6), which is now given by the kernel $K^{-1}(\varepsilon, \xi' - \xi)$, has been previously derived [10], but it is not required in present study. It should be pointed out that the meaning of the argument ξ and the parameter ε can significantly vary, e.g. an energy and related half-width ε, a position $x$ or momentum $p$ with their half-width. With regard to the problem under consideration we only require operator functions $O^1$ and $O^{-1}$ in energy space, i.e.:

$$O^{-1} = exp(0.25 \cdot \varepsilon^2 \cdot d^2/dE^2), \quad (7)$$

$$O^1 = exp(-0.25 \cdot \varepsilon^2 \cdot d^2/dE^2). \quad (7a)$$

With reference to the complete expression (2), which yields $Q^{tot}(E)$, the operators $O^1(E)$ and $O^{-1}(E)$ are the toolkit to calculate $Q^{red}(E)$. However, this situation is in fact more difficult, since we need linear combinations of the operators with different half-widths (see appendix A). Therefore, we only give an outline in this section. Since the skeleton of formula (2) consists of three Gaussian functions and one error function, the action of $O^1$ and $O^{-1}$ can be summarized by the following way:

*a. Gaussian distribution function*

$$O^{\pm 1}(\varepsilon, E) \cdot \exp(-[(E-\lambda_E)^2/\sigma^2] =$$

$$= [\sigma/(\sigma^2 \pm (-\varepsilon^2))^{1/2}] \cdot \exp(-[(E-\lambda_E)^2/(\sigma^2 \pm (-\varepsilon^2))]). \quad (8)$$

Thus $\lambda_E$ represents an arbitrary parameter (e.g. a shift or threshold energy $E_{Th}$) in formula (2), and for ε = σ the deconvolution operator $O^1(\varepsilon, E)$ converts the Gaussian distribution (8) to a δ-distribution function. Therefore we can summarize that the convolution operator $O^{-1}$ broadens a Gaussian distribution, whereas $O^1$ (deconvolution operator) affects an increased steepness of a Gaussian with decreasing half-width. It should be added that the 3 Gaussian terms appearing in $Q^{tot}(E)$ of equation (2) act with different σ-values.

*b. Error function erf(ξ)*

As the error function erf(ξ) plays a significant role with regard to the asymptotic behavior of $Q^{tot}(E)$ it appears that the statement of the complete deconvolution term resulting from the expression (2) is justified. Now the above argument ξ refers to the substitution $\xi = (E-E_{Th})/\sigma_{as}$, and the following terms result from a Lie series expansion (see appendix A):

$$O^1(\varepsilon, E) \cdot erf(\xi) = erf(\xi) + \frac{2}{\sqrt{\pi}} \cdot \sum_{n=0}^{\infty} (-1)^n \cdot$$

$$\cdot H_{2n+1}(\xi) \cdot \exp(-\xi^2) \cdot (\varepsilon/\sigma_{as})^{2n} / (n! \cdot 4^n). \quad (9)$$

The polynomials of odd order $H_{2n+1}(\xi)$ refer to as Hermite polynomials, of which the recurrence formula is stated in [10, and references therein]. The result (9) yields an interesting property, known already from quantum mechanics, namely oscillations along the asymptotic behavior of $Q^{red}(E)$, which may be regarded as characteristic resonances.

Since deconvolutions are rather sensitive and in order to avoid artifacts, we have restricted ourselves with regard to parameter fixations in equations (8) and (9). By that, $Q^{red}(E)$ may still contain 5 % - 10 % contributions of elastic and, above all, inelastic scatter of protons at nuclei. The principal difficulty comes from proper nuclear reactions changing $A_N$ or $Z$, because these processes always are closely connected with radiation transitions.

## 2. Application of deconvolution methods to $Q^{tot}(E)$ of oxygen, carbon. calcium, copper, cesium and determination of $Q^{red}(E)$

A general aspect of all applications to nuclei stated above represents the tunneling effect of protons determined previously [1] via quantum mechanical methods. This effect consists of the property that protons with kinetic energy $E_{kin} < E_{Th}$ can pass through the potential barrier (see e.g. Figure 3) and Table 1, where $E_{Th}$ is stated for the considered nuclei. However, this is only a noteworthy effect, if $E_{kin}$ of protons amounts to 90 % - 95 % of $E_{Th}$, otherwise the effect is very small and for $E_{kin} < 0.4 \cdot E_{Th}$ negligible.

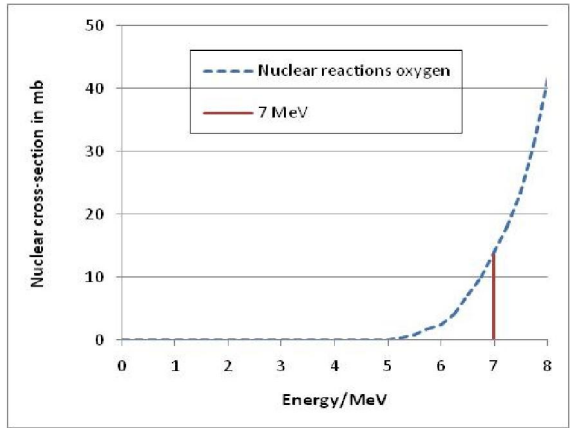

**Figure 4:** Nuclear cross-section $Q^{red}(E)$ at $E_{kin} < 8$ MeV

Figure 4 indicates that due the tunneling effect there is a small probability for a proton to surpass the potential barrier, which is 7 MeV for oxygen, but rather similar for other nuclei. Therefore Figure 4 may stand for all other nuclei considered in this study. What may happen for protons $E_{kin} < E_{Th}$? The smallest probability is that the proton can leave the strong interaction area by a further tunneling effect. Since the probability of the proton having undergone some reflections within $R_{strong}$ is extremely small to leave again the domain of strong interaction, then either *a. transitions to lower excited states with emission of γ-quanta* or *b. conversion of the proton to a neutron by exchange of a meson is possible (quantum mechanical exchange interaction due to the Pauli principle)*. The latter case b is the most probable one, since the neutron can pass through the potential wall and leave the strong interaction domain, if $E_{kin} > 0$. Then in the case of oxygen the residual nucleus is the isotope $F_9^{16}$, which undergoes ß$^+$ decay to yield again an oxygen nucleus. The behavior of the other nuclei under consideration is rather equivalent. Therefore we can summarize: If $E_{kin} < E_{Th}$, the most important reaction channel is stated for the nuclei studied in this communication (10a - 10d):

$$p + C_6^{12} \to n + N_7^{12};$$
$$\beta^+ : N_7^{12} \to C_6^{12} + e^+ + \gamma. \quad (10a)$$
$$p + O_8^{16} \to n + F_9^{16};$$
$$\beta^+ : F_9^{16} \to = O_8^{16} + e^+ + \gamma. \quad (10b)$$
$$p + Ca_{20}^{40} \to n + Sc_{21}^{40};$$
$$\beta^+ : Sc_{21}^{40} \to Ca_{20}^{40} + e^+ + \gamma. \quad (10c)$$
$$p + Cu_{29}^{63,64} \to n + Zn_{30}^{63,64};$$
$$\beta^+ : Zn_{30}^{63,64} \to Cu_{29}^{63,64} + e^+ + \gamma. \quad (10d)$$

Figure 5 can make evident some possible nuclear reaction channels indicating the importance of the impinging angle of the proton and its history by preceding scatter processes (this is referred to as Molière multiple scatter at distances $R > R_N$ or $R \gg R_N$). For this purpose, we have split the cross-section in separated zones Z1, Z2, Z3 of a spherical nuclei. As examples we look at the positions $P_1$, $P_2$, $P_3$. In order to explain Figure 5 we use the abbreviations: $E_{pot}$ represents the excitation energy of nucleon to the state of positive energy E = 0 (a neutron release with $E_{kin,n} \geq 0$ must receive the energy $E > E_{pot}$); $E_{pot,wall}$ represents the potential energy at $R_{strong}$ (this is the minimum excitation energy of a proton release, i.e., $E_{pot,wall} = E_{pot} + E_{Th}$ and $E_{kin,p} \geq E_{pot,wall}$). In the following, we shall use the abbreviations: $D_1^2$ (deuterium) and $T_1^3$ (tritium); all other nuclei and their isotopes follow the usual denominations of the periodic system.

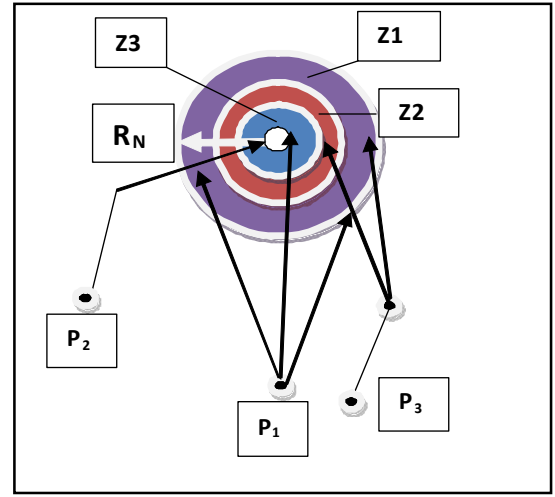

**Figure 5:** Some possible paths of protons from positions $P_1$, $P_2$, $P_3$ to hit a nucleus within different zones Z1, Z2, Z3 ($R_N$ = $R_{Nucleus}$ according to equation (1a)). The area of Z1 amounts to $\pi \cdot R_N^2 - 4 \cdot \pi \cdot R_N^2/9 = 5 \cdot \pi \cdot R_N^2/9$, the area of Z2 to $\pi \cdot R_N^2/3$, and that of Z3 to $\pi \cdot R_N^2/9$.

By regarding position $P_1$ of a proton we can verify that a nucleus can be hit in different zones and the distance from $P_1$ to a nucleus is rather far. By that, the angles between the three arrows are very small. If a nucleus exhibits approximately spherical symmetry, i.e. number of neutrons $N_n \approx Z$, the highest probability to hit the some nucleus is given by the outermost zone Z1, while it decreases significantly in direction to the central domain (Z1 → Z2 → Z3). With increasing energy of the impinging proton (E > $E_{pot}$), at first, a neutron *n* can be released. The release of *2·n, p + n, $D_1^2$, $T_1^3$* requires increasing energy again. If $N_n > Z$ (e.g. with regard to Cu) or $N_n \gg Z$ (e.g. $C_{55}^{137}$) the preference for neutron release growths. A hit at the next inner zone Z2 is only interesting for impinging protons with E $\gg$ $E_{kin,p}$ to push out nucleonic fragments (clusters) such as α-particles or an isotope of Li. Otherwise (E > $E_{pot}$) a nucleus absorbs the impinging proton and after reorganization of the nucleonic states the behavior is equivalent to the outermost zone with a preference release of neutrons. The same fact is true for a hit of the impinging proton in the central domain (Z3), where the lowest differential cross-section has to be accounted for. If E $\gg$ $E_{pot}$ a nuclear fission may occur with two fragments of about the same number of nucleons. Otherwise the behavior is comparable to the already discussed cases of the other zones,

The positions $P_2$ and $P_3$ of the impinging protons have to be associated with the same probability behavior as for $P_1$. These both positions indicate that the impinging protons may have undergone further scatter characterized by the Molière scatter theory before they enter the domain R < $R_{Nucleus}$. However, $E_{proton}$ might already be lower than for $P_1$. The consequence of the diminished energy of impinging protons may result in a reduced kinetic energy of secondary particles, and since neutrons require the lowest energy transfer for their release, the $N_n$ may increase, whereas the release of fragments growths less.

With the help of Figures 6 and 6a we are able to associate the energy for the possible nuclear reactions. Although both figures refer to the medium *'water'*, they can be transferred to other media/elements by their knowledge of $A_N$, Z and mass density ρ due to the properties of Bethe-Bloch equation [9]. However, we should like to point out that the simple substitution rule for molecules, namely Z → $Z_{sub}$ and $A_N$ → $A_{N,sub}$ is rather insufficient. Thus for water we would obtain $Z_{sub}$ = 10 and $A_{N,sub}$ = 18, which would be identical to an isotope of the noble gas $Ne_{10}^{18}$. A quantum chemical calculation of the center of mass and center of charge for water provided $Z_{sub}$ = 9.02 and $A_{N,sub}$ = 17.73 [3] with the ratio 0.5087422, whereas 10/18 yields 0,555556. This results implies that $A_N$ for water with $A_{N,sub}$ = 17.73 is a good approach due to the mass number 2·Z of oxygen, but for the effective charge $Z_{sub}$ = 10 is rather unrealistic. By use of the correct ratio for $Z_{sub}/A_{N,sub}$ for the reference system *'water'* conventional in therapy planning algorithms the transition to different media can be carried out more precisely. Taking these transitions into account we are now able to connect the zones with the corresponding areas of Bragg curves (see Figure 6) and Figure 6a yielding (determination of LET).

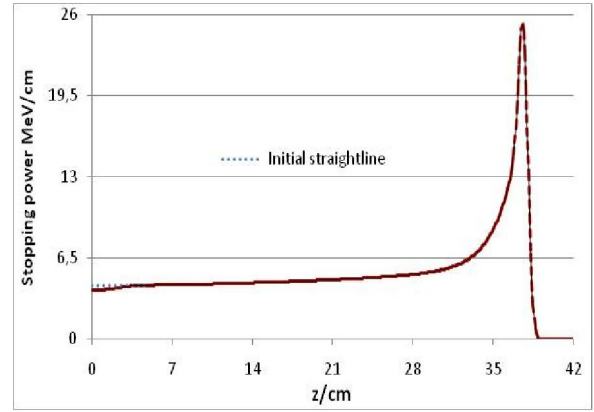

**Figure 6:** Bragg curve of 250 MeV mono-energetic protons (medium: water, solid line: calculation [9]; dashes: measurement data from M.D. Anderson synchrotron; dots: initial straight line by neglect of the Landau tail).

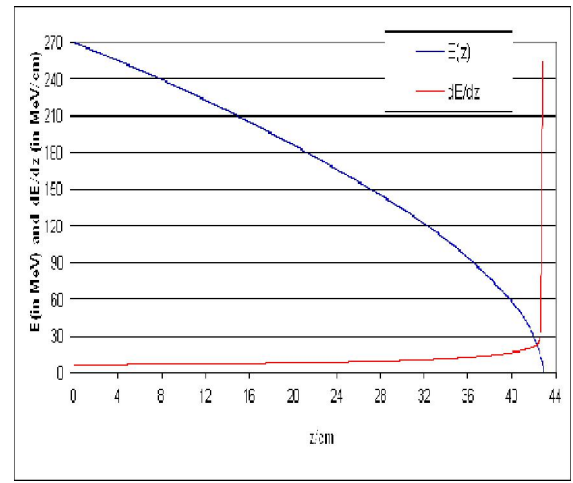

**Figure 6a:** E(z) and dE(z)/dz as a function of z (LET based on CSDA-calculation); energy straggling is omitted here in contrast to Figure 6.

Thus the very small tunneling effect with release of a neutron for $E_{proton} < E_{Th}$ occurs at the Bragg peak with growing less in direction to the distal end. The emission of a neutron, and with decreasing probability the release of *2·n, n + p, $D_1^2$, $T_1^3$* takes place in the area of exponentially ascending slope. However, all other emissions of fragment, clusters, simultaneous emissions of some single neutrons or protons with clusters and/or fragments can only occur in the initial plateau!
With regard to the zones (Figure 5) analyzing the differential cross-section we can summarize: Figures 6 and 6a serve for consecution of energy domains of specific nuclear reactions.

a. *Thus for $E \approx E_{Th}$ the proton has reached the Bragg peak and travels to distal end ($E \rightarrow 0$). Nuclear reactions in this energy domain referring to neutron release are classified by (10a - 10e) via exchange of mesons, if $R < R_{strong}$, but, in addition, $R \geq R_{Nucleus}$.*

b. *The release of nucleons (preferably neutrons) from the outermost zone Z1 of Figure 5 represent the dominant part of secondary particles in dependence of the kinetic energy $E_{proton}$ of the impinging protons.*

c. *The inner zone Z2 and the central area Z3 may also emit bigger nuclear fragments and fissions besides the release of further nucleons.*

The total nuclear cross-section $Q^{red}(E)$ resulting from Figure 5 is given by:

$$Q^{red}(E) = \iint q_{Nuclear}(E,\theta,\varphi) \cdot \sin\theta \cdot d\theta \, d\varphi. \quad (11)$$

For nuclei with spherical symmetry the integration over φ can be omitted. The total nuclear cross-section $Q^{tot}(E)$ accounts for all kinds of nuclear interactions (elastic scatter and inelastic scatter by changing some quantum numbers). The differences between $Q^{red}(E)$ and $Q^{tot}(E)$ are presented in Figures 7 - 11. In general, it can be mentioned that $Q^{red}(E)$ shows some characteristic resonance energies for nuclei with $N_n = Z$ with growing less tendency for $N_n > Z$ or $N_n \gg Z$ as true for $Cs_{55}^{137}$.

## 3. $Q^{red}(E)$ of interactions proton – oxygen/carbon/calcium/copper/cesium

Figure 7 is most important for nuclear reactions of protons in water with regard to therapy planning algorithms, since $Q^{tot}(E)$ determines the behavior of Figure 1 (decrease if fluence of primary protons), whereas Figure 8 (carbon) basicly refers to organic molecules. A general feature of all figures is that the absolute maximum of $Q^{red}(E)$ is slightly shifted to a lower energy compared to the overall maximum of $Q^{tot}(E)$.

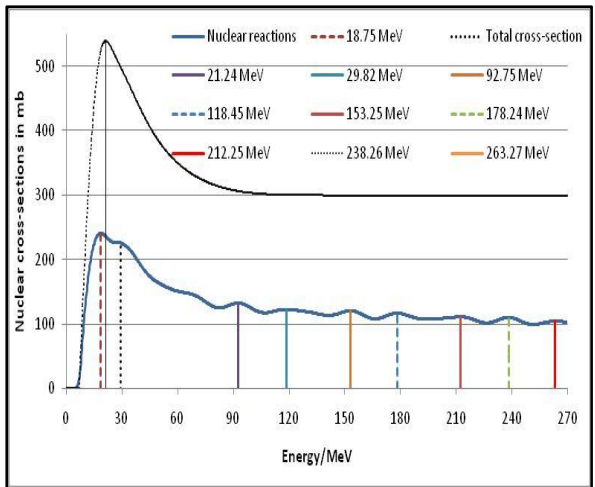

**Figure 7:** $Q^{tot}$ (dots, one maximum at 21.24 MeV)) and $Q^{red}$ (solid, 9 maxima from 18.75 MeV up to 263.27 MeV) of *oxygen*. For E > 270 MeV the existence of maxima is rather unimportant, since the asymptotic behavior is reached, and the therapeutic proton energies basicly remain below 270 MeV.

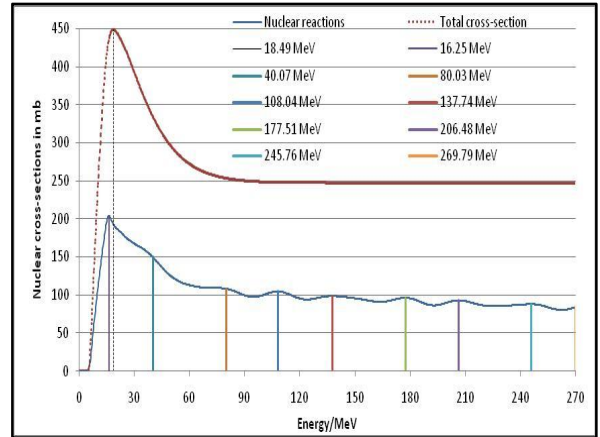

**Figure 8:** $Q^{tot}$ (dots, one maximum at 18.49 MeV) and $Q^{red}$ (solid, maxima at 16.25 MeV up 269.79 MeV) of *carbon*. If E > 270 MeV the asymptotic behavior is reached.

The nuclear cross-section $Q^{red}(E)$ of calcium (Figure 9) may rather be important in proton therapy, since the protons may either pass through bone before they can reach the given target or the irradiation of a bone tumor has to be taken into account. In particular, the branch between the first two maxima of $Q^{red}(E)$ is shifted to higher resonance energies (27.25 MeV and 46.23 MeV) indicating that the release of two neutrons or proton + neutron has been made feasible. This effect results from the property of increased number of nucleons (Z = 20 and $N_n$ = 20).

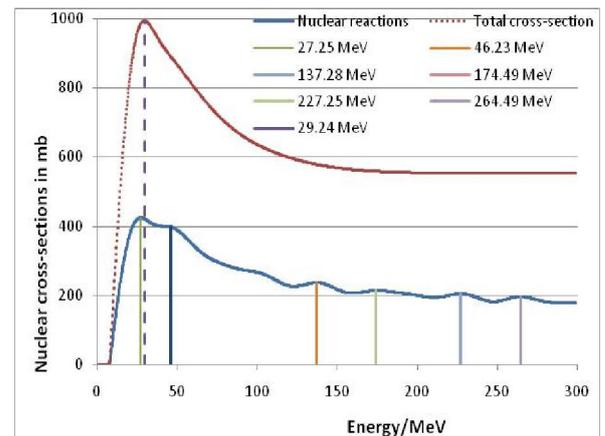

**Figure 9:** $Q^{tot}$ (dots, one maximum at 29.24 MeV) and $Q^{red}$ (solid, maxima at 27.25 MeV up to 264.49 MeV) of *calcium*.

The nuclear cross-section $Q^{red}(E)$ of Cu (Figure 10) is mainly of interest with regard to the beam-line, since the nozzle usually consists of a brass alloy of Cu and Zn. Since all nuclear properties of the Zn-nucleus are still rather equivalent to that of Cu, we do not discuss the case here for the reason of brevity.

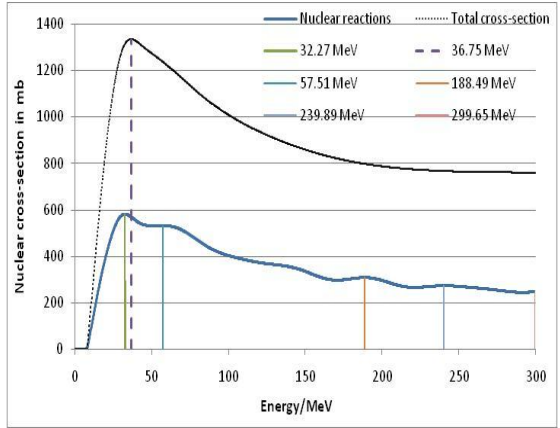

**Figure 10:** $Q^{tot}$ (dots, one maximum at 36.75 MeV) and $Q^{red}$ (solid, maxima at 32.27 MeV up to 299.65 MeV) of *copper*.

Figure 11 referring to $Cs_{55}^{137}$ certainly does not show any therapeutic relevance, since the purpose of this figure is to demonstrate that the methods presented in this study may have applications going beyond proton radiotherapy. Since it is rather known that $Cs_{55}^{137}$ belongs to the most critical reaction products of nuclear reactors due to its very long half-time of about 30 a. On the other hand, the nuclear reaction

$$p + Cs_{55}^{137} \rightarrow p + n + Cs_{55}^{136} + \gamma \quad (12)$$

leading to the isotope $Cs_{55}^{136}$ represents a rather attractive reaction product due to its half-time of ≈ 20 days. With regard to low impinging proton energies we should also like to mention the following reaction:

$$p + Cs_{55}^{137} \rightarrow n + Ba_{56}^{137} + \gamma. \quad (12a)$$

This reaction is a rather nice one, since $Ba_{56}^{137}$ is a stable nucleus.

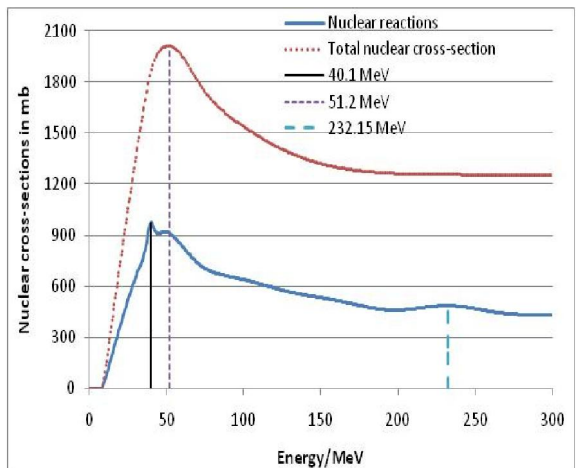

**Figure 11:** $Q^{tot}$ (dots, one maximum at 51.2 MeV) and $Q^{red}$ (solid, maxima at 40.103 MeV, 51.2 MeV, 232.15 MeV) of **cesium**.

Nuclear reactions according to the channel (12) occur in the domain between the two maxima of 40.103 and 51.2004 MeV. If we have a glance along the complete cross-section function $Q^{red}(E)$, we are able to verify that there exist various reaction processes evoking many isotopes as reaction products with Z < 55 or << 55 and $A_N$ < 137 or << 137. This possible technical application is referred to as *'transmutation'* of $Cs_{55}^{137}$ and has been taken into account as a further option besides the storage of this critical element.

## 4. Discussion and Conclusions

The behavior of secondary protons is clear: they all are able to undergo nuclear reactions with lower energies after traveling along the related medium. The reaction products (isotopes) of protons with nuclei (here oxygen) are denoted by *'heavy recoils'*, which undergo ß⁺/ß⁻ decay or electron capture with further emission of γ-quanta. However the neutrons show a rather different behavior, since their threshold energy does not exist. Some essential nuclear reactions have already previously described. Thus for the neutron energy E < 1 MeV which can be considered as 'thermal neutrons' the formation of $D_1^2$ with emission of γ-quanta is rather probable, whereas for higher energies preferably the collisions with hydrogen protons represent the main effect of neutron stopping and the formation of the oxygen isotope $O_8^{17}$ is less significant. The interaction of neutrons with environmental electrons via magnetic dipole coupling of the related spins is nearly negligible with regard to neutron stopping, since the effect is only of higher order.

With regard to Figures 7 - 11 and their maxima of the reaction channels we are able to state essential reaction types occurring in various energy domains:

$$p + X_m^{m'} \rightarrow n + X_{m+1}^{m'} + \gamma; \quad (13)$$
$$p + X_m^{m'} \rightarrow n + p + X_m^{m'-1} + \gamma; (13a)$$
$$p + X_m^{m'} \rightarrow 2 \cdot n + X_{m+1}^{m'-1} + \gamma; (13b)$$
$$p + X_m^{m'} \rightarrow D_1^2 + X_m^{m'-1} + \gamma. \quad (13c)$$

In the survey of appendix B we shall abbreviate the reaction channel (13) with $R_0$ and the channels (13a) - (13c) with $R_{1a}$, $R_{1b}$, $R_{1c}$. Thus the reaction channel (13) is only possible for lower initial proton energies (see e.g. (12a)), whereas the cases (13a) - (13c) are very probable in the outermost zone Z1 of Figure 5 and they are able to occur even at proton energies E > 200 MeV. Consequently these cases form a certain underground, when other reaction channels exhibit a maximum according to Figures 7 - 11. In the appendix B, we present the most probable nuclear reactions in dependence of the actual proton energy.

We wish finally point out two aspects referring to the role of nuclear reactions:

1. The release of neutrons preferably is of essential importance either in the domain of the Bragg peak and behind in direction of the distal end or in the area of exponential growing of the Bragg curve. Thus the algorithm of proton therapy planning systems and even Monte-Carlo calculations underrate the release of neutrons [3, 9, 12 - 15].

2. The presented method for the determination of $Q^{tot}(E)$ and $Q^{red}(E)$ only require $Z$ and $A_N$ as basic parameters. Therefore the handling of nuclear scatter processes and nuclear reactions can be carried out via voxel data based on 3D CT information using the calibrations/substitutions:

$$Z \rightarrow Z_{effective} \text{ and } A_N \rightarrow A_{N,effective}. \quad (14)$$

It should be noted that the information given by the substitution (14) is already required by handling the interaction of protons with shell electrons due to Bethe-Bloch equation [2, 3, 7, 9].

3. The reaction channels occurring with regard to $Cs_{55}^{137}$ certainly do not exhibit any therapeutic relevance, but it should be demonstrated that the calculation methods developed in this study are also applicable to technical problems such as 'transmutations' of reactor products with long half-times. In such a situation it is superfluous to deflect the proton beam before the nozzle by suitable magnets to obtain a narrow energy spectrum of the impinging protons and a rather broad energy spectrum of the proton beam is desired. Since the nuclear cross-sections $Q^{tot}(E)$ and $Q^{red}(E)$ refer to single hit processes, the yield of nuclear reactions can be strengthened by high intensity of projectile proton beams, e.g. if the first target hit leads to nucleus vibrations or excitations, and an immediate following second target hit can be able to transfer proton energy to an excited nuclear state, where the range $R_{Nuclear}$ is somewhat increased. In order to reach an optimal efficacy of the transmutations, it is necessary to perform these reactions under extremely high density concentrations with the aim to operate beyond single hit events.

## A. Appendix: Some aspects of the calculation formula (Lie series expansions and quantum mechanical background)

The Lie series expansions of the operators $O^I$ (deconvolution) and $O^{-1}$ (convolution) in the energy space are given by

$$O^{\pm 1} = \exp(\mp \frac{\varepsilon^2}{4} \cdot \frac{d^2}{dE^2}). \quad (A1)$$

$$O^1 = 1 + \sum_{n=1}^{\infty} (-1)^n \cdot \frac{\varepsilon^{2n}}{n! \cdot 4^n} \cdot \frac{d^{2n}}{dE^{2n}}. \quad (A2)$$

$$O^{-1} = 1 + \sum_{n=1}^{\infty} \frac{\varepsilon^{2n}}{n! \cdot 4^n} \cdot \frac{d^{2n}}{dE^{2n}}. \quad (A3)$$

The convolution equation of $O^{-1}$ according to (A3) is the well-known transform with a Gaussian kernel:

$$\varphi(E) = \int K(\varepsilon, E - E') \rho(E') dE'. \quad (A4)$$

The convolution integral equation of the inverse problem (A2) reads:

$$\rho(E) = \sum_{n=0}^{\infty} (-1)^n \frac{s^{2n}}{2^n \cdot n!} \int H_{2n}(\frac{E-E'}{\varepsilon}) \cdot K(\varepsilon, E - E') \varphi(E') dE'. \quad (A5)$$

With regard to the deconvolution of $Q^{tot}(E)$ providing $Q^{red}(E)$, we have to account for different values for ε, and therefore we need a linear combination of operators $O^1$. Denoting this linear combination by the operator $A^1$, which reads:

$$A^1 = c \cdot O^1(\varepsilon, E) + c_1 \cdot O_1^1(\varepsilon_1, E)$$
$$+ c_2 \cdot O_2^1(\varepsilon_2, E) + c_3 \cdot O_3^1(\varepsilon_3, E). \quad (A6)$$

The deconvolution calculation applied to $Q^{tot}(E)$ now becomes:

$$A^1 \cdot Q^{tot}(E) = Q^{red}(E). \quad (A7)$$

It should be pointed out that the inverse operator of $A^1$, namely $A^{-1}$, is defined by the relation:

$$A^1 \cdot A^{-1} = A^{-1} \cdot A^1 = 1. \quad (A8)$$

The operator $A^{-1}$ is determined via relation (A8) and the corresponding Lie series:

$$A^{-1} = \frac{1}{c \cdot O^1 + c_1 \cdot Q_1^1 + c \cdot O_2^1 + c_3 \cdot O_3^1}. \quad (A8a)$$

Thus the handling of this Lie series expansion (A8a) has previously been worked out [10]. The necessary parameters of $A^1$ are given by:

$$c = 0.4194; \; c_1 = 0.3211; \; c_2 = 0.1862;$$
$$c_3 = 0.0733; \; \varepsilon = 0.56 \cdot s_{as}; \; \varepsilon_1 = 0.41 \cdot \varepsilon;$$
$$\varepsilon_2 = 1.95 \cdot \varepsilon; \; \varepsilon_3 = 0.59 \cdot \varepsilon. \quad (A9)$$

Finally we want to go on excursion to the quantum mechanical background of the operator $O^1(E)$ and $O^{-1}(E)$. As already pointed out this formula is only defined by a Lie series expansion, which is also true in other context; e.g., formulations in the position, momentum or k-vector space. In previous considerations, we have derived in position space the corresponding form of $O^1$ by using a canonic ensemble and the Schrödinger equation. This means that we replace the momentum in the canonical ensemble by the corresponding differential operator of the momentum leading to a density operator function, which may be based on the Schrödinger equation: **p** → -i·ℏ·(∂/∂x, ∂/∂y, ∂/∂z). However, this is not possible with regard to the above formula. Therefore we have to start with a relativistic treatment. Since all physical processes with real rest mass have to occur in the time-like part of the light cone, i.e. $c^2 \cdot t^2 - x^2 > 0$, we now consider the operator formulation of the expression (in one space dimension without loss of any generality):

$$O^1 = \exp(\tfrac{c^2 \cdot t^2 - x^2}{l^2}) \rightarrow$$
$$\exp(-(\tfrac{c^2 \cdot \hbar^2}{l^2} \cdot \tfrac{\partial^2}{\partial E^2} - \tfrac{\hbar^2}{l^2} \cdot \tfrac{\partial^2}{\partial p^2})) = \exp[-(\tfrac{c^2 \cdot \hbar^2}{l^2} \cdot \tfrac{\partial^2}{\partial E^2} + C)]. \quad (A10)$$

The right-hand side of the above equation follows from the differential operator representation of the well-known canonical commutation rules *[x, p] = iℏ* and *[E, t] = iℏ*. The phase factor function $\exp(i \cdot p \cdot l / \hbar)$ referred to as C can be omitted with regard to calculations in the energy space; it results from differentiations in the momentum space:

$$\exp(\tfrac{\hbar^2}{l^2} \cdot \tfrac{\partial^2}{\partial p^2}) \cdot \exp(\tfrac{i \cdot p \cdot l}{\hbar}) = \exp(\tfrac{i \cdot p \cdot l}{\hbar} - 1). \quad (A10a)$$

We are preferably interested in the influence of the energy in the time-like part of the light cone, whereas by taking account of the momentum would provide this influence, which may enrich the formalism, if the momentum is stated in 3 dimensions. The operator $O^{-1}$ consequently is space-like. This fact has not been an object of the above considerations. In connection with above methods we use the abbreviation:

$$\varepsilon = \tfrac{2 \cdot \hbar \cdot c}{l}. \quad (A11)$$

Then by neglecting the momentum contribution we finally obtain the formula introduced above, which we have used for partial deconvolution of $Q^{tot}(E)$ according to sections 2 and 3. The length *'l'* along the light cone can be determined by experimental data used in the deconvolution procedure or by Monte-Carlo calculations. It is also possible to ignore ε according to relation (A11) and to use as a free optimization parameter without any quantum mechanical background.

## B. Appendix: Nuclear reactions channels in dependence of the actual proton energy and the role of resonances in Figures 7 - 11

Owing to the deconvolution procedure applied in this study we have been able to separate the energy transfer of the projectile proton to elastic/inelastic scatter and that part being in connection with nuclear reaction channels. The above mentioned figures indicate that the number of maxima decreases with increasing Z and $A_N$, and, above all, the property $N_n > Z$ and $N_n \gg Z$ seems to be decisive. Thus the cross-sections according to Figures 7 and 8 are important in irradiation of soft tissue targets. Figure 9 is important, when either a proton passes bone tissue or the target is incorporated by a bone. Figures 10 and 11 basically are of interest in technical applications (beam-guide of proton beams and transmutations). A general feature of rather low proton energies, i.e. $E_{prot} < E_{pot}$, is the exchange interaction between proton and a nucleon (denoted by X) due to meson exchange with a particular role of Pauli principle of the overall system:

$$p + X_m^{m'} : \left| \begin{array}{c} proton \Leftrightarrow nucleus \\ meson \; exchange \end{array} \right\rangle \rightarrow n + X_{m+1}^{m'}. \quad (B1)$$

The resulting 'heavy recoil' $X_{m+1}^{m'}$ can either undergo a ß$^+$ decay (10a - 10d) or become immediately a stable isotope as true in the case (12a). This reaction type always occurs with regard to every nucleus, if the available proton energy is $E_{prot} < E_{pot,wall}$ for protons and $E_{prot} < E_{pot}$ for neutrons. In Figure 7 and 8 the first maxima ($C_6^{12}$ at 16.25 MeV and $O_8^{16}$ at 18.75 MeV) are exclusively determined by (B1), whereas for $Ca_{20}^{40}$ $Cu_{29}^{63,64}$, 36.75 MeV) and Figure 11 ($Cs_{55}^{137}$, 40.1 MeV) the first maxima include further reactions. Therefore only the first maximum of carbon related to (B1) is peculiar, since a second maximum of $Q^{red}$ in the immediate environ does not exist and $Q^{red}$ continuously is growing down to 80.03 MeV. This behavior results from specific properties of this nucleus with the smallest $R_{Nucleus}$ considered here favoring higher fractions of elastic/inelastic scatter. The very small second maximum of $O_8^{16}$ (29.82 MeV) enables the reactions (13a) and (13b). All (Figure 9, 27.25 MeV) and the cases of Figure 10 ( other cases (Figures 9 - 11) contain the reactions within the first maxima (13a) and (13b), and the second maximum also includes the reaction (13c). In the following, we use the normalization value *'1'* with regard to the weight of all possible nuclear reactions up to proton energies of 270 MeV/300 MeV, and we recall that $R_0$ refers to the channel (13) and $R_{1a}$, $R_{1b}$, $R_{1c}$ to (13a - 13c). The types $R_{1a,b,c}$ occur with regard to the total energy domain, and we attribute them rather higher relevance.

Tables 3 - 7 present the nuclear reaction channels of the reaction p + X → secondary particles + *heavy recoils* and their possible decay products depending of the actual proton energy $E_{proton}$ ($E_p$), which is stated within boundaries. If the weight of a channel fraction is ≤ 0.001, then it is ignored and neighboring fraction is corresponding rounded or added to a suitable channel; the extremely small fractions of $R_0$ and $R_1$ in Z3 is added to Z1. The specific weight of each channel is stated in the related parenthesis, if more than one channel is possible. In the following tables we also shall use the abbreviations:

$p + X_m^{m'} \to D_1^2 + p + X_{m-1}^{m'-2} + \gamma$ ($R_{1d}$); $p + X_m^{m'} \to D_1^2 + n + X_m^{m'-2} + \gamma$ ($R_{1e}$); $p + X_m^{m'} \to T_1^3 + X_m^{m'-2} + \gamma$ ($R_{2a}$); $p + X_m^{m'} \to T_1^3 + p + X_{m-1}^{m'-3} + \gamma$ ($R_{2b}$); $p + X_m^{m'} \to T_1^3 + n + X_m^{m'-3} + \gamma$ ($R_{2c}$); $p + X_m^{m'} \to \alpha + X_{m-1}^{m'-3} + \gamma$ ($R_{2d}$); $p + X_m^{m'} \to \alpha + n + X_{m-1}^{m'-4} + \gamma$ ($R_{2e}$); $p + X_m^{m'} \to \alpha + p + n + X_{m-2}^{m'-5} + \gamma$ ($R_{2f}$); $p + X_m^{m'} \to \alpha + 2n + X_{m-1}^{m'-5} + \gamma$ ($R_{2g}$); $p + X_m^{m'} \to \alpha + p + 2n + X_{m-2}^{m'-6} + \gamma$ ($R_{2h}$); $p + X_m^{m'} \to Li_3^5 + X_{m-2}^{m'-4} + \gamma$ ($R_{3a}$); $p + X_m^{m'} \to Li_3^6 + X_{m-2}^{m'-5} + \gamma$ ($R_{3b}$); $p + X_m^{m'} \to Li_3^7 + X_{m-2}^{m'-6} + \gamma$ ($R_{3c}$); $p + X_m^{m'} \to Li_3^5 + n + X_{m-2}^{m'-5} + \gamma$ ($R_{3d}$); $p + X_m^{m'} \to Be_4^7 + X_{m-3}^{m'-6} + \gamma$ ($R_{4a}$); $p + X_m^{m'} \to Be_4^8 + X_{m-3}^{m'-7} + \gamma$ ($R_{4b}$); $p + X_m^{m'} \to Be_4^9 + X_{m-3}^{m'-8} + \gamma$ ($R_{4c}$).

**Table 3:** List of nuclear reactions of the proton - oxygen interaction with *m= 8 and m' =16* yielding $p + O_8^{16} \to X_{m-q}^{m'-r}$ + *secondary particles* and their weights. The overall weights of the zones Z1, Z2. Z3 amount to Z1: w = 0.56; Z2: w = 0.33; Z3: w = 0.11.

| Z1: $E_{proton}$/MeV | | |
|---|---|---|
| lower/upper boundaries | $w_1$ | channel type |
| 0 > $E_p$ ≤ 22.95 | 0.08 | $R_0$ |
| 22.95 > $E_p$ ≤ 65.03 | 0.07 | $R_0$ (0.02), $R_{1a}$ (0.02), $R_{1b}$ (0.02), $R_{1c}$ (0.01) |
| 65.03 > $E_p$ ≤ 105.17 | 0.07 | $R_0$ (0.01), $R_{1a}$ (0.02), $R_{1b}$ (0.02), $R_{1c}$ (0.01), $R_{1d}$ (0.01) |
| 105.17 > $E_p$ ≤ 138.21 | 0.07 | $R_{1a}$ (0.01), $R_{1b}$ (0.01), $R_{1c}$ (0.01), $R_{1d}$ (0.02), $R_{1e}$ (0.02) |
| 138.21 > $E_p$ ≤ 176.34 | 0.06 | $R_{1c}$ (0.01), $R_{1d}$ (0.01), $R_{1e}$ (0.01), $R_{2a}$ (0.03) |
| 176.34 > $E_p$ ≤ 191.58 | 0.05 | $R_{1e}$ (0.01), $R_{2a}$ (0.02), $R_{2b}$ (0.02) |
| 191.58 > $E_p$ ≤ 206.72 | 0.04 | $R_{1e}$ (0.01), $R_{2a}$ (0.01), $R_{2b}$ (0.01), $R_{2c}$ (0.01) |
| 206.72 > $E_p$ ≤ 222.30 | 0.04 | $R_{2a}$ (0.01), $R_{2b}$ (0.01), $R_{2c}$ (0.02) |
| 235.64 > $E_p$ ≤ 254.92 | 0.04 | $R_{2a}$ (0.01), $R_{2b}$ (0.01), $R_{2c}$ (0.01), $R_{2d}$ (0.01) |
| 254.92 > $E_p$ ≤ 270.00 | 0.04 | $R_{2b}$ (0.01), $R_{2c}$ (0.01), $R_{2d}$ (0.01), $R_{2e}$ (0.01) |
| Z2: $E_{proton}$/MeV | | |
| lower/upper boundaries | $w_2$ | channel type |
| 0 > $E_p$ ≤ 22.95 | - | - |
| 22.95 > $E_p$ ≤ 65.03 | 0.05 | $R_{1a}$ (0.01), $R_{1b}$ (0.02), $R_{1c}$ (0.02) |
| 65.03 > $E_p$ ≤ 105.17 | 0.05 | $R_{1b}$ (0.01), $R_{1c}$ (0.02), $R_{1e}$ (0.02) |
| 105.17 > $E_p$ ≤ 138.21 | 0.05 | $R_{1b}$ (0.01), $R_{1c}$ (0.01), $R_{2a}$ (0.01), $R_{2b}$ (0.02) |
| 138.21 > $E_p$ ≤ 176.34 | 0.05 | $R_{2a}$ (0.01), $R_{2b}$ (0.01), $R_{2c}$ (0.01), $R_{2d}$ (0.02) |
| 176.34 > $E_p$ ≤ 191.58 | 0.04 | $R_{2b}$ (0.01), $R_{2c}$ (0.01), $R_{2d}$ (0.01), $R_{2e}$ (0.01) |
| 191.58 > $E_p$ ≤ 206.72 | 0.03 | $R_{2b}$ (0.005), $R_{2c}$ (0.005), $R_{2d}$ (0.005), $R_{2e}$ (0.005), $R_{2f}$ (0.01) |
| 206.72 > $E_p$ ≤ 222.30 | 0.02 | $R_{2b}$ (0.004), $R_{2c}$ (0.004), $R_{2d}$ (0.004), $R_{2e}$ (0.004), $R_{2f}$ (0.002), $R_{2g}$ (0.002) |
| 235.64 > $E_p$ ≤ 254.92 | 0.02 | $R_{2b}$ (0.002), $R_{2c}$ (0.002), $R_{2d}$ (0.003), $R_{2e}$ (0.003), $R_{2f}$ (0.004), $R_{2g}$ (0.004), $R_{2h}$ (0.002) |
| 254.92 > $E_p$ ≤ 270.00 | 0.02 | $R_{2b}$ (0.002), $R_{2c}$ (0.002), $R_{2d}$ (0.002), $R_{2e}$ (0.002), $R_{2f}$ (0.003), $R_{2g}$ (0.003), $R_{2h}$ (0.003). $R_{3a}$ (0.003) |
| Z3: $E_{proton}$/MeV | | |
| lower/upper boundaries | $w_3$ | channel type |
| 0 > $E_p$ ≤ 22.95 | - | - |
| 22.95 > $E_p$ ≤ 65.03 | 0.010 | $R_{1b}$ (0.004), $R_{1c}$ (0.004), $R_{1d}$ (0.002) |
| 65.03 > $E_p$ ≤ 105.17 | 0.010 | $R_{1b}$ (0.003), $R_{1c}$ (0.003), $R_{1d}$ (0.002), $R_{1e}$ (0.002) |
| 105.17 > $E_p$ ≤ 138.21 | 0.010 | $R_{1b}$ (0.002), $R_{1c}$ (0.002), $R_{1d}$ (0.002), $R_{1e}$ (0.002), $R_{2a}$ (0.002) |
| 138.21 > $E_p$ ≤ 176.34 | 0.010 | $R_{1d}$ (0.002), $R_{1e}$ (0.002), $R_{2a}$ (0.002), $R_{2b}$ (0.002), $R_{2c}$ (0.002) |
| 176.34 > $E_p$ ≤ 191.58 | 0.010 | $R_{2d}$ (0.002), $R_{2e}$ (0.002), $R_{2f}$ (0.002), $R_{2g}$ (0.002), $R_{2h}$ (0.002) |
| 191.58 > $E_p$ ≤ 206.72 | 0.010 | $R_{2d}$ (0.001), $R_{2e}$ (0.001), $R_{2f}$ (0.001), $R_{2g}$ (0.001), $R_{2h}$ (0.002), $R_{3a}$ (0.002), $R_{3b}$ (0.001), $R_{3c}$ (0.001) |
| 206.72 > $E_p$ ≤ 222.30 | 0.015 | $R_{2g}$ (0.002), $R_{2h}$ (0.002), $R_{3a}$ (0.002), $R_{3b}$ (0.003), $R_{3c}$ (0.003), $R_{3d}$ (0.003) |
| 235.64 > $E_p$ ≤ 254.92 | 0.015 | $R_{3a}$ (0.002), $R_{3b}$ (0.002), $R_{3c}$ (0.002), $R_{3d}$ (0.003), $R_{4a}$ (0.003), $R_{4b}$ (0.003) |
| 254.92 > $E_p$ ≤ 270.00 | 0.020 | $R_{3a}$ (0.003), $R_{3b}$ (0.003), $R_{3c}$ (0.003), $R_{3d}$ (0.003), $R_{4a}$ (0.003), $R_{4b}$ (0.003), $R_{4c}$ (0.002) |

As already pointed out, the heavy recoils of oxygen ($O_{8-q}^{16-r}$) usually undergo ß$^+$-decay; a rather important case is the decay reaction (10b) $F_9^{16} \rightarrow O_8^{16} + e_0^+$ with $T_{1/2} = 22$ sec. With regard to the recoils of oxygen, all decay processes are rather fast (i.e. less than 10 minutes). Since the heavy recoils stated Table 3, their decay products and half-times can be taken from web, we do not intend to present this subject of matter here. The noteworthy neutron release may also imply reactions of neutrons with oxygen, and an important case is the channel:

$$n + O_8^{16} \rightarrow p + N_7^{16} + \gamma; \; (\beta^- : N_7^{16} \rightarrow O_8^{16} + e_0^- + \gamma, T_{1/2} = 120 \; \text{sec}). \quad (B2)$$

Reactions of the type (B2) occur by many modifications in those heave recoils resulting from nuclei with $Z < N_n$, e.g. $Cu_{29}^{63}$ or $Cs_{55}^{137}$.

**Table 4:** List of nuclear reactions of the proton - carbon interaction with $m = 6$ and $m' = 12$ yielding $p + C_6^{12} \rightarrow X_{6-q}^{12-r}$ + *secondary particles*, weights and decay products. The overall weights of the zones Z1, Z2. Z3 amount to Z1: $w_1 = 0.56$; Z2: $w_2 = 0.33$; Z3: $w_3 = 0.11$.

| Z1: $E_{proton}$/MeV lower/upper boundaries | $w_1$ | channel type |
|---|---|---|
| $0 > E_p / \leq 20.92$ | 0.10 | $R_0$ |
| $20.92 > E_p \leq 42.47$ | 0.09 | $R_0$ (0.03), $R_{1a}$ (0.03), $R_{1b}$ (0.02), $R_{1c}$ (0.01) |
| $42.47 > E_p \leq 92.65$ | 0.08 | $R_0$ (0.02), $R_{1a}$ (0.02), $R_{1b}$ (0.02), $R_{1c}$ (0.01), $R_{1d}$ (0.01) |
| $92.65 > E_p / \leq 122.28$ | 0.07 | $R_{1a}$ (0.02), $R_{1b}$ (0.02), $R_{1c}$ (0.01), $R_{1d}$ (0.01), $R_{1e}$ (0.01) |
| $122.28 > E_p \leq 164.17$ | 0.06 | $R_{1c}$ (0.02), $R_{1d}$ (0.02), $R_{1e}$ (0.01), $R_{2a}$ (0.01) |
| $164.17 > E_p \leq 192.53$ | 0.05 | $R_{1e}$ (0.01), $R_{2a}$ (0.02), $R_{2b}$ (0.01), $R_{2c}$ (0.01) |
| $192.53 > E_p \leq 224.65$ | 0.04 | $R_{1e}$ (0.01), $R_{2a}$ (0.01), $R_{2b}$ (0.01), $R_{2c}$ (0.01) |
| $224.65 > E_p \leq 255.37$ | 0.04 | $R_{2b}$ (0.01), $R_{2c}$ (0.01), $R_{2d}$ (0.01), $R_{2e}$ (0.01) |
| $255.37 > E_p \leq 270.00$ | 0.03 | $R_{2b}$ (0.01), $R_{2c}$ (0.01), $R_{2d}$ (0.005), $R_{2e}$ (0.005) |
| **Z2: $E_{proton}$/MeV lower/upper boundaries** | $w_2$ | **channel type** |
| $0 > E_p \leq 20.92$ | - | - |
| $20.92 > E_p \leq 42.47$ | 0.05 | $R_0$ (0.01), $R_{1a}$ (0.01), $R_{1b}$ (0.015), $R_{1c}$ (0.015) |
| $42.47 > E_p \leq 92.65$ | 0.04 | $R_{1b}$ (0.01), $R_{1c}$ (0.02), $R_{1d}$ (0.01) |
| $92.65 > E_p \leq 122.28$ | 0.04 | $R_{1b}$ (0.01), $R_{1c}$ (0.01), $R_{1d}$ (0.01), $R_{1e}$ (0.01) |
| $122.28 > E_p \leq 164.17$ | 0.04 | $R_{2a}$ (0.01), $R_{2b}$ (0.01), $R_{2c}$ (0.01), $R_{2d}$ (0.01) |
| $164.17 > E_p \leq 192.53$ | 0.04 | $R_{2b}$ (0.005), $R_{2c}$ (0.005), $R_{2d}$ (0.01), $R_{2e}$ (0.02) |
| $192.53 > E_p \leq 224.65$ | 0.04 | $R_{2b}$ (0.005), $R_{2c}$ (0.005), $R_{2d}$ (0.005), $R_{2e}$ (0.005), $R_{2f}$ (0.01), $R_{2g}$ (0.01) |
| $224.65 > E_p \leq 255.37$ | 0.04 | $R_{2b}$ (0.005), $R_{2c}$ (0.005), $R_{2d}$ (0.005), $R_{2e}$ (0.005), $R_{2f}$ (0.005), $R_{2g}$ (0.005), $R_{2h}$ (0.01) |
| $255.37 > E_p \leq 270.00$ | 0.04 | $R_{2b}$ (0.004), $R_{2c}$ (0.004), $R_{2d}$ (0.004), $R_{2e}$ (0.004), $R_{2f}$ (0.004), $R_{2g}$ (0.005), $R_{2h}$ (0.005). $R_{3a}$ (0.01) |
| **Z3: $E_{proton}$/MeV lower/upper boundaries** | $w_3$ | **channel type** |
| $0 > E_p \leq 20.92$ | - | - |
| $20.92 > E_p \leq 42.47$ | 0.015 | $R_{1b}$ (0.003), $R_{1c}$ (0.003), $R_{1d}$ (0.003), $R_{1e}$ (0.003), $R_{2a}$ (0.003) |
| $42.47 > E_p / \leq 92.65$ | 0.015 | $R_{1b}$ (0.003), $R_{1c}$ (0.003), $R_{1d}$ (0.003), $R_{1e}$ (0.002), $R_{2a}$ (0.002), $R_{2b}$ (0.002) |
| $92.65 > E_p \leq 122.28$ | 0.015 | $R_{1b}$ (0.002), $R_{1c}$ (0.002), $R_{1d}$ (0.002), $R_{1e}$ (0.003), $R_{2a}$ (0.002), $R_{2b}$ (0.002), $R_{2c}$ (0.002) |
| $122.28 > E_p / \leq 164.17$ | 0.015 | $R_{1d}$ (0.002), $R_{1e}$ (0.002), $R_{2a}$ (0.002), $R_{2b}$ (0.003), $R_{2c}$ (0.003), $R_{2d}$ (0.002) |
| $164.17 > E_p \leq 192.53$ | 0.015 | $R_{2d}$ (0.002), $R_{2e}$ (0.002), $R_{2f}$ (0.002), $R_{2g}$ (0.003), $R_{2h}$ (0.003), $R_{3a}$ (0.002) |
| $192.53 > E_p / \leq 224.65$ | 0.015 | $R_{2g}$ (0.002), $R_{2h}$ (0.002), $R_{3a}$ (0.002), $R_{3b}$ (0.003), $R_{3c}$ (0.003), $R_{3d}$ (0.002) |
| $224.65 > E_p \leq 255.37$ | 0.01 | $R_{3a}$ (0.002), $R_{3b}$ (0.002), $R_{3c}$ (0.003), $R_{3d}$ (0.003) |
| $255.37 > E_p \leq 270.00$ | 0.01 | $R_{3a}$ (0.0015), $R_{3b}$ (0.0015), $R_{3c}$ (0.003), $R_{3d}$ (0.004) |

With regard to the following nuclei with an increased number of Z and, above all, $N_n$ we need additional abbreviations for nuclear reaction channels in order to be able to represent all additional possibilities by a reasonable way. The purpose of the reaction channels ($R_{5a}$) to ($R_{6k}$) is to account for fissions of the nuclei $Ca_{20}^{40}$, $Cu_{29}^{63}$ and $Cs_{55}^{137}$ to two nuclei with about the half of Z and the half of $N_n$. With regard to the comparably light nuclei $C_6^{12}$ and $O_8^{16}$ this problem is already accounted for by considering the isotopes of *lithium* and *oxygen* according to the reaction channels ($R_{3a}$) - ($R_{4d}$):

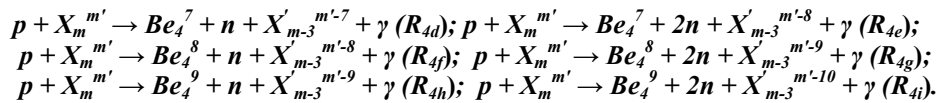

$p + X_m^{m'} \rightarrow Be_4^7 + n + X'_{m-3}^{m'-7} + \gamma \; (R_{4d}); \; p + X_m^{m'} \rightarrow Be_4^7 + 2n + X'_{m-3}^{m'-8} + \gamma \; (R_{4e});$
$p + X_m^{m'} \rightarrow Be_4^8 + n + X'_{m-3}^{m'-8} + \gamma \; (R_{4f}); \; p + X_m^{m'} \rightarrow Be_4^8 + 2n + X'_{m-3}^{m'-9} + \gamma \; (R_{4g});$
$p + X_m^{m'} \rightarrow Be_4^9 + n + X'_{m-3}^{m'-9} + \gamma \; (R_{4h}); \; p + X_m^{m'} \rightarrow Be_4^9 + 2n + X'_{m-3}^{m'-10} + \gamma \; (R_{4i}).$

**If m and m' even ($Ca_{20}^{40}$):**

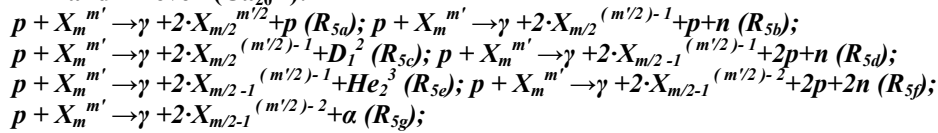

$p + X_m^{m'} \rightarrow \gamma + 2 \cdot X_{m/2}^{m'/2} + p \; (R_{5a}); \; p + X_m^{m'} \rightarrow \gamma + 2 \cdot X_{m/2}^{(m'/2)-1} + p + n \; (R_{5b});$
$p + X_m^{m'} \rightarrow \gamma + 2 \cdot X_{m/2}^{(m'/2)-1} + D_1^2 \; (R_{5c}); \; p + X_m^{m'} \rightarrow \gamma + 2 \cdot X_{m/2-1}^{(m'/2)-1} + 2p + n \; (R_{5d});$
$p + X_m^{m'} \rightarrow \gamma + 2 \cdot X_{m/2-1}^{(m'/2)-1} + He_2^3 \; (R_{5e}); \; p + X_m^{m'} \rightarrow \gamma + 2 \cdot X_{m/2-1}^{(m'/2)-2} + 2p + 2n \; (R_{5f});$
$p + X_m^{m'} \rightarrow \gamma + 2 \cdot X_{m/2-1}^{(m'/2)-2} + \alpha \; (R_{5g});$

**If m and m' odd ($Cu_{29}^{63}$, $Cs_{55}^{137}$):**

$p + X_m^{m'} \rightarrow \gamma + p + X_{(m-1)2}^{(m'-1)/2} + X_{(m+1)2}^{(m'+1)/2}$ ($R_{6a}$); $p + X_m^{m'} \rightarrow \gamma + p + n + X_{(m-1)2}^{m'/2} + X_{(m+1)2}^{m'/2}$ ($R_{6b}$);

$p + X_m^{m'} \rightarrow \gamma + D_1^2 + X_{(m-1)2}^{m'/2} + X_{(m+1)2}^{m'/2}$ ($R_{6c}$); $p + X_m^{m'} \rightarrow \gamma + 2p + n + X_{(m-1)2}^{m'/2} + X_{(m-1)2}^{m'/2}$ ($R_{6d}$);

$p + X_m^{m'} \rightarrow \gamma + He_2^3 + X_{(m-1)2}^{m'/2} + X_{(m-1)2}^{m'/2}$ ($R_{6e}$); $p + X_m^{m'} \rightarrow \gamma + 2p + 2n + X_{(m-1)2}^{(m'-1)/2} + X_{(m-1)2}^{(m'-1)/2}$ ($R_{6f}$);

$p + X_m^{m'} \rightarrow \gamma + \alpha + X_{(m-1)2}^{(m'-1)/2} + X_{(m-1)2}^{(m'-1)/2}$ ($R_{6g}$); $p + X_m^{m'} \rightarrow \gamma + 2p + 3n + X_{(m-1)2}^{(m'-1)/2} + X_{(m-1)2}^{(m'-3)/2}$ ($R_{6h}$);

$p + X_m^{m'} \rightarrow \gamma + \alpha + n + X_{(m-1)2}^{(m'-1)/2} + X_{(m-1)2}^{(m'-3)/2}$ ($R_{6i}$); $p + X_m^{m'} \rightarrow \gamma + 2p + 4n + X_{(m-1)2}^{(m'-3)/2} + X_{(m-1)2}^{(m'-3)/2}$ ($R_{6j}$);

$p + X_m^{m'} \rightarrow \gamma + \alpha + 2n + X_{(m-1)2}^{(m'-3)/2} + X_{(m-1)2}^{(m'-3)/2}$ ($R_{6k}$);

One has to be aware of that the sequences ($R_{5a}$) - ($R_{6d}$) may suitably be continued. However, we shall account for higher order terms by adding them to the nearest neighbor.

**Table 5:** List of nuclear reactions of the proton - calcium interaction with *m = 20 and m' = 40* yielding $p + Ca_{20}^{40} \rightarrow X_{m-q}^{m'-r}$ + *secondary particles*, weights and decay products. The overall weights of the zones Z1, Z2, Z3 amount to Z1: $w_1 = 0.56$; Z2: $w_2 = 0.33$; Z3: $w_3 = 0.11$.

| Z1: $E_{proton}$/MeV lower/upper boundaries | $w_1$ | channel type |
|---|---|---|
| $0 > E_p \leq 30.85$ | 0.10 | $R_0$ (0.08), $R_{1a}$ (0.02) |
| $30.85 > E_p \leq 68.13$ | 0.10 | $R_{1a}$ (0.02), $R_{1b}$ (0.02), $R_{1c}$ (0.02), $R_{1d}$ (0.02), $R_{1e}$ (0.02) |
| $68.13 > E_p \leq 155.39$ | 0.10 | $R_{1c}$ (0.01), $R_{1d}$ (0.01), $R_{1e}$ (0.02), $R_{2a}$ (0.02), $R_{2b}$ (0.02), $R_{2c}$ (0.02) |
| $155.39 > E_p \leq 199.64$ | 0.09 | $R_{2a}$ (0.01), $R_{2b}$ (0.01), $R_{2c}$ (0.01), $R_{2d}$ (0.01), $R_{2e}$ (0.01), $R_{2f}$ (0.02), $R_{2g}$ (0.01), $R_{2h}$ (0.01) |
| $199.64 > E_p \leq 242.03$ | 0.09 | $R_{2c}$ (0.01), $R_{2d}$ (0.01), $R_{2e}$ (0.01), $R_{2f}$ (0.01), $R_{2g}$ (0.01), $R_{2h}$ (0.01), $R_{3a}$ (0.02), $R_{3b}$ (0.01) |
| $242.03 > E_p \leq 300.00$ | 0.08 | $R_{2d}$ (0.01), $R_{2e}$ (0.01), $R_{2f}$ (0.01), $R_{2g}$ (0.01), $R_{2h}$ (0.01), $R_{3a}$ (0.01), $R_{3b}$ (0.01), $R_{3c}$ (0.01) |
| **Z2: $E_{proton}$/MeV** lower/upper boundaries | $w_2$ | channel type |
| $0 > E_p / \leq 30.85$ | - | - |
| $30.85 > E_p \leq 68.13$ | 0.08 | $R_{1b}$ (0.01), $R_{1c}$ (0.01), $R_{1d}$ (0.01), $R_{1e}$ (0.01), $R_{2a}$ (0.01), $R_{2b}$ (0.01), $R_{2c}$ (0.02) |
| $68.13 > E_p \leq 155.39$ | 0.08 | $R_{2c}$ (0.01), $R_{2d}$ (0.01), $R_{2e}$ (0.01), $R_{2f}$ (0.01), $R_{2g}$ (0.01), $R_{2h}$ (0.01), $R_{3a}$ (0.01), $R_{3b}$ (0.01) |
| $155.39 > E_p \leq 199.64$ | 0.07 | $R_{3a}$ (0.01), $R_{3b}$ (0.01), $R_{3c}$ (0.01), $R_{3d}$ (0.01), $R_{4a}$ (0.01), $R_{4b}$ (0.01), $R_{4c}$ (0.01) |
| $199.64 > E_p \leq 242.03$ | 0.06 | $R_{4b}$ (0.01), $R_{4c}$ (0.01), $R_{4d}$ (0.01), $R_{4e}$ (0.01), $R_{4f}$ (0.005), $R_{4g}$ (0.005), $R_{4h}$ (0.005), $R_{4i}$ (0.005) |
| $242.03 > E_p \leq 300.00$ | 0.04 | $R_{5a}$ (0.01), $R_{5b}$ (0.005), $R_{5c}$ (0.005), $R_{5d}$ (0.005), $R_{5e}$ (0.005), $R_{5f}$ (0.005), $R_{5g}$ (0.005) |
| **Z3: $E_{proton}$/MeV** lower/upper boundaries | $w_3$ | channel type |
| $0 > E_p / \leq 30.85$ | - | - |
| $30.85 > E_p \leq 68.13$ | 0.020 | $R_{2a}$ (0.002), $R_{2b}$ (0.002), $R_{2c}$ (0.002), $R_{2d}$ (0.002), $R_{2e}$ (0.002), $R_{2f}$ (0.003), $R_{2g}$ (0.003), $R_{2h}$ (0.004) |
| $68.13 > E_p \leq 155.39$ | 0.015 | $R_{3a}$ (0.003), $R_{3b}$ (0.003), $R_{3c}$ (0.002), $R_{3d}$ (0.002), $R_{4a}$ (0.02), $R_{4b}$ (0.001), $R_{4c}$ (0.001), $R_{4d}$ (0.001) |
| $155.39 > E_p / \leq 199.64$ | 0.015 | $R_{4e}$ (0.001), $R_{4f}$ (0.001), $R_{4g}$ (0.001), $R_{4h}$ (0.001), $R_{4i}$ (0.001), $R_{5a}$ (0.004), $R_{5b}$ (0.003), $R_{5c}$ (0.003) |
| $199.64 > E_p \leq 242.03$ | 0.015 | $R_{5a}$ (0.001), $R_{5b}$ (0.001), $R_{5c}$ (0.001), $R_{5d}$ (0.003), $R_{5e}$ (0.003), $R_{5f}$ (0.002), $R_{5g}$ (0.002), $R_{5i}$ (0.002) |
| $242.03 > E_p \leq 300.00$ | 0.015 | $R_{5a}$ (0.001), $R_{5b}$ (0.001), $R_{5c}$ (0.001), $R_{5d}$ (0.001), $R_{5e}$ (0.001), $R_{5f}$ (0.001), $R_{5g}$ (0.001), $R_{5i}$ (0.002) $R_{5j}$ (0.003), $R_{5k}$ (0.003) |

**Table 6:** List of nuclear reactions of the proton - copper interaction with *m = 29 and m' = 63* yielding $p + Cu_{29}^{63} \rightarrow X_{m-q}^{m'-r}$ + *secondary particles*, weights and decay products. The overall weights of the zones Z1, Z2. Z3 amount to Z1: $w_1 = 0.56$; Z2: $w_2 = 0.33$; Z3: $w_3 = 0.11$.

| Z1: $E_{proton}$/MeV lower/upper boundaries | $w_1$ | channel type |
|---|---|---|
| $0 > E_p \leq 49.47$ | 0.15 | $R_0$ (0.1), $R_{1a-1e}$ (0.05) |
| $49.47 > E_p \leq 155.53$ | 0.14 | $R_{1a-1e}$ (0.03), $R_{2a-2h}$ (0.06), $R_{3a-3d}$ (0.06) |
| $155.53 > E_p \leq 219.62$ | 0.14 | $R_{3a-3d}$ (0.05), $R_{4a-4i}$ (0.09) |
| $219.62 > E_p \leq 300.00$ | 0.13 | $R_{3a-3d}$ (0.01), $R_{4a-4i}$ (0.12) |
| **Z2: $E_{proton}$/MeV** lower/upper boundaries | $w_2$ | channel type |
| $0 > E_p / \leq 49.47$ | 0.03 | $R_0$ (0.01), $R_{1a-1e}$ (0.02) |
| $49.47 > E_p \leq 155.53$ | 0.11 | $R_{1a-1e}$ (0.03), $R_{2a-2h}$ (0.04), $R_{3a-3d}$ (0.04) |
| $155.53 > / \leq 219.62$ | 0.1 | $R_{3a-3d}$ (0.02), $R_{4a-4i}$ (0.08) |
| $219.62 > E_p \leq 300.00$ | 0.09 | $R_{4a-4i}$ (0.09) |
| **Z3: $E_{proton}$/MeV** lower/upper boundaries | $w_3$ | channel type |
| $0 > E_p \leq 49.47$ | - | - |
| $49.47 > E_p \leq 155.53$ | 0.04 | $R_{3a-3d}$ (0.04) |
| $155.53 > E_p \leq 219.62$ | 0.04 | $R_{4a-4i}$ (0.04) |
| $219.62 > E_p \leq 300.00$ | 0.03 | $R_{6a-6k}$ (0.03) |

**Table 7:** List of nuclear reactions of the proton - cesium interaction with *m = 55 and m' = 137* yielding ***p + Cs$_{55}^{137}$ → X$_{m-q}^{m'-r}$ + secondary particles***, weights and decay products. The overall weights of the zones Z1, Z2. Z3 amount to Z1: $w_1 = 0.56$; Z2: $w_2 = 0.33$; Z3: $w_3 = 0.11$.

| Z1: E$_{proton}$/MeV lower/upper boundaries | $w_1$ | channel type |
|---|---|---|
| 0 > E$_p$ ≤ 51.20 | 0.20 | R$_0$ (0.12), R$_{1a-1e}$ (0.08) |
| 51.20 > E$_p$ ≤ 198.47 | 0.18 | R$_{1a-1e}$ (0.02), R$_{2a-2h}$ (0.04), R$_{3a-3d}$ (0.06), R$_{4a-4i}$ (0.06) |
| 198.47 > E$_p$ ≤ 300.00 | 0.18 | R$_{3a-3d}$ (0.02), R$_{4a-4i}$ (0.16) |
| **Z2: E$_{proton}$/MeV lower/upper boundaries** | **$w_2$** | **channel type** |
| 0 > E$_p$ ≤ 51.20 | - | - |
| 51.20 > E$_p$ ≤ 198.47 | 0.17 | R$_{1a-1e}$ (0.02), R$_{2a-2h}$ (0.04), R$_{3a-3d}$ (0.04), R$_{4a-4i}$ (0.07) |
| 198.47 > E$_p$ ≤ 300.00 | 0.16 | R$_{4a-4i}$ (0.10), R$_{6a-6k}$ (0.06) |
| **Z3: E$_{proton}$/MeV lower/upper boundaries** | **$w_3$** | **channel type** |
| 0 > E$_p$ ≤ 51.20 | - | |
| 51.20 > E$_p$ ≤ 198.47 | 0.05 | R$_{1a-1e}$ (0.005), R$_{2a-2h}$ (0.015), R$_{3a-3d}$ (0.01), R$_{4a-4i}$ (0.02) |
| 198.47 > E$_p$ ≤ 300.00 | 0.06 | R$_{6a-6k}$ (0.06) |

With reference to the overall weight coefficients *w* of the zones Z1, Z2, Z3 we have to form $w_1 \cdot \pi \cdot R_N^2$, $w_2 \cdot \pi \cdot R_N^2$ and $w_3 \cdot \pi \cdot R_N^2$ in order to obtain the real weights of the nuclear reactions. If the subscripts in tables 6 and 7 start with the corresponding index and finish with the highest index, then all contributions incorporate approximately the identical weight factor, e.g. R$_{1a-1e}$, R$_{2a-2h}$, etc.

Some final notes in the appendix B seem to be justified:

1. It is evident and has already been mentioned that the release of neutrons and, by that, possible reaction channels evoked these neutrons need considerable improvement in therapy planning systems. This is particularly evident due to the high RBE of neutrons.

2. Radiation effects occurring in connection with proton irradiation are not sufficiently accounted for. There three origins of radiation creation: Interaction of protons with nuclei by creating γ-quanta, emission of γ-quanta of the heavy recoil nuclei, and ß$^+$-decay of these nuclei yielding to production of γ-quanta via annihilation of positrons by collisions with electrons. These photons lead to a broader lateral scatter of dose profiles and reduce the LET and RBE in the environment of the Bragg peak.